\documentclass[aps,prd,floats,superscriptaddress,altaffilletter]{revtex4}
\usepackage{amsfonts,amsmath,amssymb}
\usepackage{graphicx}
\newcommand{\sgn}{\operatorname{{\mathrm sgn}}}
\begin{document}

\newcommand{\be}{\begin{equation}}
\newcommand{\ee}{\end{equation}}
\newcommand{\bea}{\begin{eqnarray}}
\newcommand{\eea}{\end{eqnarray}}
\def \wc {w_{c,\text{eff}}}
\def \wv {w_{\varphi,\text{eff}}}

\title{Dynamics of dark energy with a coupling to dark matter}

\author{Christian G. B\"ohmer}
\email{c.boehmer@ucl.ac.uk} \affiliation{Department of
Mathematics, University College London, London WC1E 6BT, UK}

\author{Gabriela Caldera-Cabral}
\email{gaby.calderacabral@port.ac.uk} \affiliation{Institute of
Cosmology \& Gravitation, University of Portsmouth, Portsmouth PO1
2EG, UK}

\author{Ruth Lazkoz}
\email{ruth.lazkoz@ehu.es} \affiliation{Fisika Teorikoa, Euskal
Herriko Unibertsitatea, 48080 Bilbao, Spain}

\author{Roy Maartens}
\email{roy.maartens@port.ac.uk} \affiliation{Institute of
Cosmology \& Gravitation, University of Portsmouth, Portsmouth PO1
2EG, UK}

\date{\today}

\begin{abstract}

Dark energy and dark matter are the dominant sources in the
evolution of the late universe. They are currently only indirectly
detected via their gravitational effects, and there could be a
coupling between them without violating observational constraints.
We investigate the background dynamics when dark energy is
modelled as exponential quintessence, and is coupled to dark
matter via simple models of energy exchange. We introduce a new
form of dark sector coupling, which leads to a more complicated
dynamical phase space and has a better physical motivation than
previous mathematically similar couplings.

\end{abstract}

\maketitle

\section{Introduction}

Observations are providing increasingly compelling evidence that
the expansion of the Universe is accelerating, driven by ``dark
energy" (see e.g.~\cite{Spergel:2006hy} for recent results). The
simplest model of dark energy is a cosmological constant
$\Lambda$, representing the vacuum energy density, and this model
provides a very good fit to a range of independent observations.
However, there is no satisfactory theoretical explanation for the
very small value of $\Lambda$. Furthermore, the $\Lambda$ model
suffers from a fine-tuning, or ``coincidence", problem -- why is
the dark matter density comparable to the vacuum energy density
now, given that their time evolution is so different?

If the dark energy evolves with time, this may alleviate the
coincidence problem. The simplest models of evolving dark energy
are light scalar fields, known as ``quintessence". If the
quintessence is coupled to the dark matter, then this may be able
to account for the similar energy densities in the dark sector
today. A decisive way of achieving similar energy densities is if
the coupling leads to an accelerated scaling attractor solution,
with
 \begin{equation} \label{scal}
{\Omega_{\text{dark energy}}\over \Omega_{\text{dark
matter}}}=O(1) ~~\mbox{ and }~~ \ddot a >0\,.
 \end{equation}
In this case, the coincidence problem is reduced to a simple
choice of parameters to match $ \Omega_{\text{dark energy}}/
\Omega_{\text{dark matter}}$ to observations. Since the
accelerated scaling solution is an attractor, no fine-tuning of
initial conditions is needed. However, the dynamics that produces
such scaling in the dark sector may have other undesirable
consequences.

Here we study quintessence with an exponential potential,
\begin{equation}
V(\varphi)= V_0 \exp \left(-\kappa \lambda  \varphi\right)\,,~~
\kappa^2:=8\pi G\,,
\end{equation}
where $\lambda$ is dimensionless and $V_0>0$. The dynamics of a
universe with exponential quintessence and an uncoupled perfect
fluid have been studied, and these models do not admit late-time
accelerated scaling attractors that satisfy
Eq.~(\ref{scal})~\cite{quin}. When we introduce a coupling between
the quintessence and dark matter, accelerated scaling attractors
are possible~\cite{Wetterich:1994bg,Amendola:1999qq}. However, in
some models, this is achieved at the expense of introducing other
problems which can rule out the model~\cite{Amendola:2006qi}.

A general coupling between a quintessence field $\varphi$ and dark
matter (with density $\rho_c$) may be described in the background
by the balance equations,
\begin{eqnarray}
  \label{relation}
\dot \rho_c  &=& - 3H\rho_c-Q\,,\label{cc}\\
\dot \rho_{\varphi} &=& - 3H(1+w_{\varphi})\rho_{\varphi}+ Q\,.
\label{kg1}
\end{eqnarray}
Here $Q$ is the rate of energy density exchange in the dark
sector, and
 \be \label{sq}
Q~\left\{ \begin{array}{l} >0\\ <0 \end{array} \right. ~~
\Rightarrow ~~ \mbox{energy transfer}~\left\{ \begin{array}{l}
\mbox{dark matter $\to$ dark energy}\\ \mbox{dark energy $\to$
dark matter}\end{array} \right.
 \ee
The dark energy equation of state parameter is
 \be
w_\varphi:={ p_\varphi \over \rho_\varphi}={{1\over
2}\dot\varphi^2-V(\varphi) \over {1\over
2}\dot\varphi^2+V(\varphi)}\,.\label{wp}
 \ee
The modified Klein-Gordon equation follows from Eq.~(\ref{kg1}):
\begin{equation}
\ddot \varphi +3H \dot \varphi + \frac{dV}{d\varphi}= {Q \over
\dot\varphi}\,.
  \label{kg}
\end{equation}
When we include baryons ($ \rho_b$) and radiation ($\rho_r $), the
remaining evolution equations are
\begin{eqnarray}
\dot \rho_b &=& - 3H\rho_b \,,\\
\dot {\rho}_r &=& - 4H\rho_r \,,\\
\dot H  &=&  -\frac{\kappa^2}{2}\left[\rho_c+\rho_b+
\frac{4}{3}\rho_r+\dot\varphi^2 \right], \label{ray}
  \end{eqnarray}
subject to the Friedman constraint,
\begin{eqnarray}
  \label{fried}
\Omega_c+\Omega_b+\Omega_r
+\Omega_{\varphi}=1\,,~~ \Omega := \frac{\kappa^2
\rho}{3H^2}.
\end{eqnarray}

We can define effective equation of state parameters for the dark
sector, which describe the equivalent uncoupled model in the
background: $\dot \rho_c +3H(1+ \wc)\rho_c=0 $, $\dot \rho_\varphi
+3H(1+ \wv)\rho_\varphi=0 $. By Eqs.~(\ref{cc}) and (\ref{kg1}),
 \be \label{weff}
\wc={Q \over 3H\rho_c}\,,~~ \wv= w_\varphi -{Q \over 3H
\rho_\varphi}\,.
 \ee
It follows that
 \bea
&& Q >0 ~~ \Rightarrow ~~ \left\{
\begin{array}{ll} \wc>0 & ~~\mbox{dark matter redshifts faster
than}~a^{-3}\\ \wv<w_\varphi & ~~\mbox{dark energy has more
accelerating power}\end{array} \right.
\\
&& Q <0 ~~ \Rightarrow ~~ \left\{
\begin{array}{ll}\wc<0 & ~~\mbox{dark matter redshifts slower
than}~a^{-3}\\
\wv>w_\varphi & ~~\mbox{dark energy has less accelerating power}
\end{array} \right.
 \eea
When $Q>0$ it is possible that $\wv<-1$ (see~\cite{Huey:2004qv}
for specific examples). This means that the coupled quintessence
behaves like a ``phantom" uncoupled model -- but without any
negative kinetic energies.

Equations~(\ref{cc})--(\ref{ray}) are an autonomous system of the
form
\begin{eqnarray}
 \dot {\mathbf{x}} = \mathbf{f}(\mathbf{x})\,,
  \label{math1}
\end{eqnarray}
and the critical points satisfy $\mathbf{f}(\mathbf{x}_*)=0$. In
order to study the stability of the critical points, we expand
about them, $\mathbf{x} = \mathbf{x}_* +\mathbf{u}$, and
Eq.~(\ref{math1}) yields
\begin{equation}
\dot {\mathbf{u}} = {\mathbf{f}'}(\mathbf{x}_*) \mathbf{u} +
\mathbf{g}(\mathbf{x})\,.
\end{equation}
Here $\mathbf{g}(\mathbf{x})/||\mathbf{x}|| \rightarrow 0$ as
$\mathbf{x} \rightarrow \mathbf{x}_*$, and
\begin{equation}\label{eigv}
f'_{ij}(\mathbf{x}_*) = \frac{\partial f_{i}}{\partial
x_{j}}(\mathbf{x}_*)\,,
\end{equation}
is a constant non-singular matrix, whose eigenvalues encode the
behaviour of the dynamical system near the critical point.

If a component of $\mathbf{f}$ can be written as a fraction
$u(\mathbf{x})/v(\mathbf{x})$, then a critical point requires the
vanishing of the numerator, $u(\mathbf{x}_*)=0$. If the
denominator also vanishes at the critical point,
$v(\mathbf{x}_*)=0$, then care is needed in obtaining the
eigenvalues of the linearized system~(\ref{math1}). Strictly
speaking, the fraction $u(\mathbf{x}_*)/v(\mathbf{x}_*)$ may not
be well defined. However, it is still possible to obtain
analytical results via analysis of the behavior of the eigenvalues
of $\mathbf{f}'$ in the limit $v(\mathbf{x}) \rightarrow 0$.

\section{Models of the dark sector coupling}
\label{review}

There is as yet no basis in fundamental theory for a specific
coupling in the dark sector, and therefore any coupling model will
necessarily be phenomenological, although some models will have a
more physical justification than others. Various models of energy
exchange have been considered. Some of these are simple functional
ansatzes, such as $Q \propto a^n$. However these models are
incomplete: they cannot be thoroughly tested against observations,
since one has no idea what the perturbation of $Q$ should be.

A satisfactory model requires at least that $Q$ should be
expressed in terms of the energy densities and other covariant
quantities. Two simple examples represent two of the main types of
model:
 \bea
\mbox{(I) } && \quad\quad Q=\sqrt{ 2 /3}\, \kappa\,\beta
\rho_c\dot\varphi\,,
\label{I} \\
\mbox{(II)} && \quad\quad Q= \alpha H \rho_c\,,\label{II}
 \eea
where $\beta$ and $\alpha$ are dimensionless constants whose sign
determines the direction of energy transfer, according to
Eq.~(\ref{sq}):
 \be
\alpha,\beta~\left\{ \begin{array}{l} >0\\ <0 \end{array} \right.
~~ \Rightarrow ~~ \mbox{energy transfer}~\left\{ \begin{array}{l}
\mbox{dark matter $\to$ dark energy}\\ \mbox{dark energy $\to$
dark matter}\end{array} \right.
 \ee
For model (II), the case $\alpha>0$ corresponds to the decay of
dark matter into dark energy, with decay rate $ \alpha H$.

Coupling (I) may be motivated within the context of scalar-tensor
theories~\cite{Wetterich:1994bg,Amendola:1999qq,Holden:1999hm}.
Generalizations of this model allow for $\beta=\beta(\varphi)$ and
more general forms of $V(\varphi)$ (see, e.g.,
Refs.~\cite{Huey:2004qv,Amendola:2003wa}). Couplings which generalize or are closely related to model
(II) have been considered as well, see, e.g.,
Refs.~\cite{Zimdahl:2001ar} for $Q/H=\alpha_c \rho_c+\alpha_x \rho_x$ and  Ref.~\cite{Guo:2007zk} for $Q/H=\alpha \Omega_x$.

For simplicity, we neglect the baryons (which are not coupled to
dark energy), and we neglect radiation (since we are mainly
interested in the late universe). The Friedman
constraint~(\ref{fried}) becomes
 \be  \label{constr}
\Omega_c+\Omega_\varphi = 1\,,
 \ee
and the total equation of state parameter is given by
 \be\label{tot}
w_{\text{tot}}:= {p_{\text{tot}} \over
\rho_{\text{tot}}}={p_\varphi \over
\rho_\varphi+\rho_c}=w_\varphi\Omega_\varphi ~~\mbox{and}~~
\dot\rho_{\text{tot}} + 3H(1+w_{\text{tot}})\rho_{\text{tot}}=0\,.
 \ee
The condition for acceleration is $w_{\text{tot}}<-1/3$.

We introduce the same dimensionless variables $x,y$ as in the
uncoupled case~\cite{quin}, where
\begin{eqnarray}
  x^2=\frac{\kappa^2 \dot {\varphi}^2}{{6}H^2}\,,~~
  y^2=\frac{\kappa^2 {V}}{{3}H^2}\,,
  \label{def1}
\end{eqnarray}
and Eq.~(\ref{constr}) implies that
\begin{equation}
0\leq\Omega_\varphi=x^2+y^2 \leq 1\,.
\end{equation}
In the new variables, the equation of state parameters are
 \be
w_\varphi={x^2-y^2 \over x^2+y^2}\,, ~~ w_{\text{tot}} =
x^2-y^2\,.
 \ee
At a critical point $(x_*,y_*)$, it follows that
 \be
a(t) \propto t^{2/3(1+x_*^2-y_*^2)}~~\mbox{and}~~ \ddot
a>0~\mbox{if}~ x_*^2-y_*^2<-{1\over 3} \,.
 \ee

The Hubble evolution equation may be written as
\begin{equation}
  \frac{ \dot H}{H^2}=- \frac{3}{2}(1+x^2-y^2)\,.
  \label{reH}
\end{equation}
The energy balance equations~(\ref{cc}) and (\ref{kg1}) for
coupling models (I) and (II) are independent of $H$ when expressed
in the variables $x(N)$ and $y(N)$, where $N=\ln a$. Thus
Eq.~(\ref{reH}) is not needed for these coupling models, and the
phase space is two-dimensional $(x,y)$ space.

\subsection*{Coupling Model (I): $~~Q=\sqrt{ 2 /3}\, \kappa\,\beta
\rho_c\dot\varphi$}

The autonomous system is
\begin{eqnarray}
  \label{x1}
x' & = & -3x+\lambda {\sqrt{6}\over 2}\, y^2+
\frac{3}{2}x(1+x^2-y^2)
+ \beta(1-x^2-y^2)\,,\\
  \label{y1}
y' & = & -\lambda {\sqrt{6}\over 2}\,
xy+\frac{3}{2}y(1+x^2-y^2)\,,
\end{eqnarray}
where a prime denotes $ d/d N$. The critical points and the
conditions for stability, acceleration ($w_{\text{tot} *}<-1/3$)
and physical existence, are summarized in Table~\ref{crit}, where
for convenience we have introduced the parameters
 \be\label{bb}
b:=\lambda -{\sqrt{6}\over 3}\, \beta\,,~~ B_\pm= {\sqrt{6}\,b\,
[\, \pm (b^2-3/2)^{3/2}-b(b^2-39/8) ] \over 4b^2+3/4}\,.
 \ee
Table~\ref{crit} combines all the possible cases ($\lambda$ and
$\beta$ negative and positive), and recovers the particular
results of previous
work~\cite{Amendola:1999qq,Billyard:2000bh,Holden:1999hm} in the
case of pressure-free matter ($w_c=0$).

\begin{table*}[!h]
\begin{center}
\begin{tabular}
{|@{\hspace{0.03in}}c@{\hspace{0.03in}}|@{\hspace{0.03in}}c
@{\hspace{0.03in}}|
@{\hspace{0.03in}}c@{\hspace{0.03in}}|@{\hspace{0.03in}}c
@{\hspace{0.03in}}| @{\hspace{0.03in}}c@{\hspace{0.03in}}|
@{\hspace{0.03in}}c@{\hspace{0.03in}}|@{\hspace{0.03in}}c
@{\hspace{0.03in}}| @{\hspace{0.03in}}c@{\hspace{0.05in}}|} \hline
Point& $x_*$ & $y_*$ & Stable? & $\Omega_{\varphi *}$  &
$w_{\text{tot} *}$&Acceleration? & Existence?
\\
[0.1cm] \hline
\hline &&&&&&&\\[-0.3cm]
A &1 &0 &  $\beta>\displaystyle{3\over2}$ and $\lambda> \sqrt{6}$
& 1&
 1 & no&  all  $\lambda,\beta$ \\[0.2cm]
\hline&&&&&&&\\[-0.3cm]
B &$-1$ &0 &  $\beta<-\displaystyle{3\over2}$ and $\lambda<-
\sqrt{6}$ &
1 & 1 & no&  all  $\lambda,\beta$\\
[0.2cm]
\hline&&&&&&&\\[-0.3cm]
C & $\displaystyle\frac{\lambda}{\sqrt{6}}$ &
$\displaystyle\sqrt{1-\frac{\lambda^2}{6}}$ & $0<\lambda^2 <6$ and
& 1 &$\displaystyle\frac{\lambda^2}{3}-1 $ & $\lambda^2<2$&
$\lambda^2\leq 6$ \\&&& $\lambda^2-\displaystyle {\beta \over \sqrt{6}}
\lambda-3<0$;&&&&\\
   &    & &  $\lambda=0$ & & & &
\\[0.2cm]
\hline&&&&&&&\\[-0.3cm]
D &$\displaystyle\frac{\sqrt{6}}{{2}b }$&
$\displaystyle\frac{\sqrt{9-2 \sqrt{6} \beta  b }}{\sqrt{6} b }$ &
$\displaystyle b^2\geq {3\over 2}$ and: &
$\displaystyle\frac{9-\sqrt{6} \beta b }{3 b ^2}$&
$\displaystyle\frac{\sqrt{6} \beta }{3 b} $&$\displaystyle {\beta
\over b}<- {\sqrt{6} \over 6}$
&  $\displaystyle b^2\geq {3\over 2}$ and\\
&&&$B_{\text{sgn}(b)}<\beta b < \displaystyle{3\sqrt{6} \over 4} $
or &&& & $\displaystyle \frac{\sqrt{6}}{2}(3-b^2 ) \leq \beta b
\leq \frac{3\sqrt{6}}{4}$
\\&&&
$\displaystyle \frac{\sqrt{6}}{2}(3-b^2 )< \beta b
<B_{-\text{sgn}(b)} $&&&&
\\[0.4cm]
\hline&&&&&&&\\[-0.3cm]
E &$\displaystyle\frac{2\beta}{3}$ &0 &   stable when it exists
 & $\displaystyle\frac{4\beta^2}{9}$
 &$\displaystyle\frac{4\beta^2}{9}$ & no &
 $\displaystyle\beta \leq \frac{3}{2}$\\ &&&and is hyperbolic
&&&&
\\[0.2cm]
\hline
\end{tabular}
\end{center}
\caption[crit]{\label{crit} The properties of the critical points
for the coupling model (I). Here $b$ and $B_\pm $ are defined in
Eq.~(\ref{bb}).}
\end{table*}

\begin{figure*}[!ht]
\centering
\includegraphics[width=0.48\textwidth]{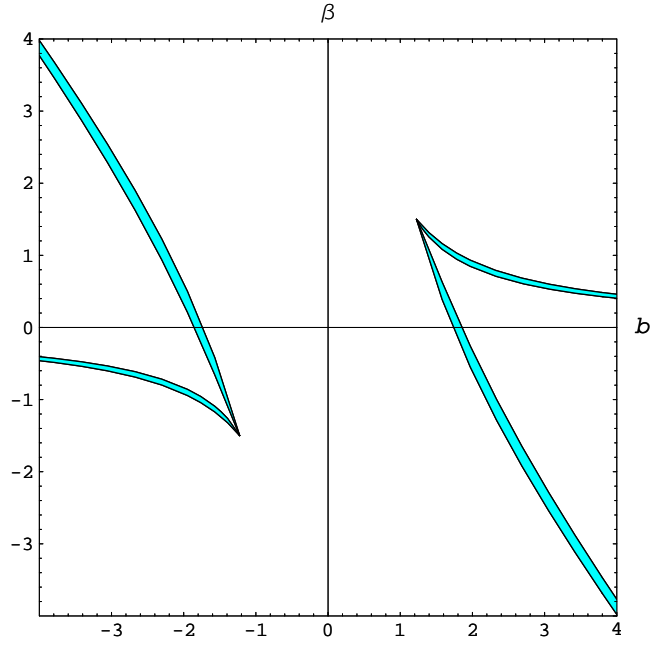}
\caption{The region of stability (blue, shaded) in the $(b,\beta$)
parameter space, for critical point D in coupling model (I). }
\label{DI}
\end{figure*}

The complicated stability conditions for critical point D were
confirmed numerically, and the results are shown in Fig.~\ref{DI}.
This point is an accelerated scaling solution that allows for
Eq.~(\ref{scal}) to be satisfied, i.e.,
 \begin{equation} \label{scal2}
0<\Omega_{c *}\,,\Omega_{\varphi *}<1 ~\mbox{  and  }~
w_{\text{tot} *} <-{1\over 3}\,.
 \end{equation}
Acceleration requires $\beta/b<-\sqrt{6}/6$. This includes
positive and negative $\beta$ (provided that $\lambda$ is
accordingly restricted), but in the uncoupled case, $\beta=0$,
acceleration is not possible for the scaling solution.

Point E is also a scaling solution, but it is always decelerating.

\subsection*{Coupling Model (II):  $~~Q=\alpha H \rho_c$}

The autonomous system of equations in this case is
\begin{eqnarray}
  \label{x2}
x' & = & -3x + \lambda {\sqrt{6}\over 2}\,y^2 +
\frac{3}{2} x(1+x^2-y^2) + {\alpha}\,{(1-x^2-y^2) \over 2x}\,,\\
  \label{y2}
y' & = & -\lambda {\sqrt{6}\over 2}\,xy +
\frac{3}{2}y(1+x^2-y^2)\,.
\end{eqnarray}
We summarize the critical points and their properties in
Table~\ref{crit2}. Our results cover all signs of $\lambda$ and
$\alpha$ and agree with previous results~\cite{Billyard:2000bh}
for $\lambda<0, \alpha<0$. Two additional critical points, E and
F, occur for $\alpha>0$.

\begin{table*}[!ht]
\begin{center}
\begin{tabular}{|c|c|c|c|c|c|c|c|} \hline
Point& $x_*$ & $y_*$ & Stable? & $\Omega_{\varphi *}$  &
$w_{\text{tot} *}$& Acceleration? &
Existence? \\
\hline \hline A & 1 &0 & $\alpha>3$ and $\lambda>\sqrt{6}$   & 1
&1& no&  all $\lambda,\alpha$  \\
\hline B &-1& 0 &  $\alpha>3$ and $\lambda<-\sqrt{6}$   & 1 & 1
&no&  all $\lambda,\alpha$
\\
\hline C &$\displaystyle\frac{\lambda}{\sqrt{6}}$ &$
\sqrt{1-\displaystyle\frac{\lambda^2}{6}} $ & $\alpha>\lambda^2-3$
and $ \lambda^2 <6$ & 1 & $\displaystyle\frac{\lambda^2}{3}-1$&
$\lambda^2<2$ & $\lambda^2 \leq 6$
\\
\hline D & $\displaystyle\frac{\alpha+3}{\sqrt{6}\lambda}$ &
$\displaystyle\frac{\sqrt{(\alpha+3)^2-2\alpha\lambda^2}}
{\sqrt{6}\lambda}$ & see Fig.~\ref{DII}
&$\displaystyle\frac{(\alpha+3)^2-\alpha\lambda^2}{3\lambda^2} $ &
$\displaystyle\frac{\alpha }{3} $ & $\alpha<-1$ &$\alpha^2 \leq 9$
and
\\&&&&&&&
$2\alpha \leq \displaystyle{(\alpha+3)^2 \over \lambda^2} \leq
\alpha +3$
\\
\hline E & $\displaystyle\frac{\sqrt{\alpha}}{\sqrt{3}}$ & $0$ &
$\lambda>\sqrt{6}$ and & $\displaystyle\frac{\alpha}{3}$ &
$\displaystyle\frac{\alpha}{3}$
 & no &  $0\leq \alpha\leq 3$ \\
   &    &        & $C<\alpha<3$      &  & & &
\\
\hline F &$-\displaystyle\frac{\sqrt{\alpha}}{\sqrt{3}}$  &$0$ &
$\lambda<-\sqrt{6}$ and & $\displaystyle\frac{\alpha}{3}$  &
$\displaystyle\frac{\alpha}{3}$ & no&  $0\leq \alpha\leq 3$
\\   &    &        & $C<\alpha<3$      &  & & &
\\
\hline
\end{tabular}
\end{center}
\caption[crit2]{\label{crit2} The properties of the critical
points for the coupling model (II). Here $C:= \lambda^2-3 +
\sqrt{\lambda^2(\lambda^2+6)}$. }
\end{table*}

\begin{figure*}[!ht]
\centering
\includegraphics[width=0.48\textwidth]{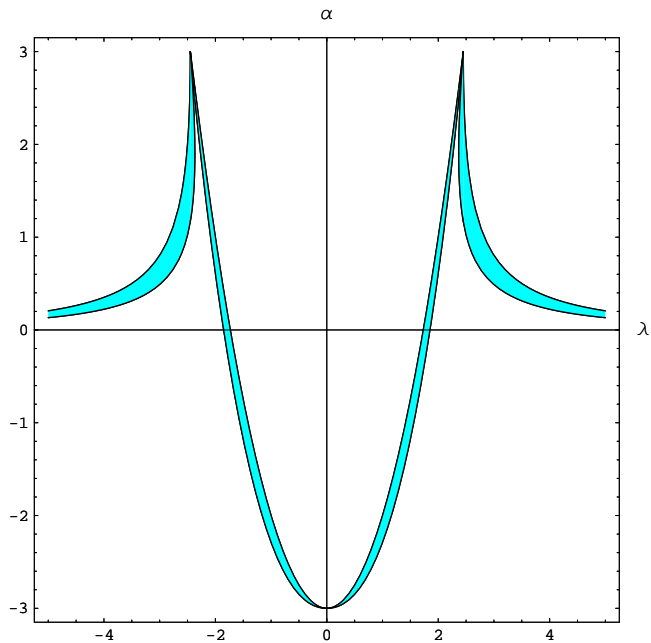}
\caption{The region of stability (blue, shaded) in the
$(\lambda,\alpha$) parameter space, for critical point D in
coupling model (II). } \label{DII}
\end{figure*}

Furthermore, we have performed numerical integrations to probe the
complicated stability conditions for the critical point D. The
results are summarized in Fig.~\ref{DII}. Point D again allows for
accelerated scaling solutions that satisfy Eq.~(\ref{scal}),
provided that $\alpha<-1$. In the uncoupled case $\alpha=0$, and
for positive $\alpha$ (decay of dark matter into dark energy),
acceleration is not possible for D.

Points E and F are also scaling solutions, but they are always
decelerating.

\section{Coupling model (III): $~~Q=\Gamma \rho_c$}
\label{curvaton}

Coupling model (I) was introduced via scalar-tensor
theory~\cite{Wetterich:1994bg,Amendola:1999qq}, and was shown to
produce accelerated scaling solutions. Although model (I) has a
clear physical motivation, it is ruled out by
observations~\cite{Amendola:2006qi}: the accelerated scaling
attractor is not connected to a matter era where structure grows
in the standard way. Indeed, generalizations of (I) with
$\beta=\beta(\varphi)$ face the same
problem~\cite{Amendola:2006qi}.

Coupling model (II) is not based on a physical model of dark
sector interactions, but is designed for mathematical simplicity.
This model and its simple generalization~\cite{Zimdahl:2001ar},
$Q= \alpha H(\rho_c+ \rho_x)$, are specifically designed to
produce an accelerated scaling attractor. Model (II) and its
generalization evade the problems that model (I) has with a
non-standard matter era~\cite{Olivares:2007rt}. These models are
useful for phenomenology, but it is difficult to see how they can
emerge from a physical description of dark sector interactions.
The rate of transfer, $\alpha H$, is determined by the expansion
rate of the universe -- rather than by purely local quantities
associated with particle/ field interactions.

In order to avoid this problem, we follow~\cite{val} and replace
the non-local transfer rate of model (II) by a local rate
$\Gamma$,
 \be
\mbox{(III)} \quad\quad Q=\Gamma \rho_c\,,\label{III}
 \ee
where we assume that $\Gamma$ is constant. This form of $Q$ is
used
in other contexts. When $\Gamma>0$, the same $Q$ is used in:\\
(1)~a simple model to describe the decay of dark matter into
radiation~\cite{Cen:2000xv}, \\(2)~a simple model for the decay of
a curvaton field into radiation~\cite{Malik:2002jb},
\\ (3)~a special case of a model in which superheavy dark matter
particles decay to a quintessence scalar
field~\cite{Ziaeepour:2003qs}.

When $\Gamma>0$, the energy transfer in Eq.~(\ref{III})
corresponds to the decay of dark matter into dark energy. Models
with decay of dark matter to dark energy allow for the possibility
that there is no dark energy field in the very early universe, and
that dark energy ``condenses" as a result of the slow decay of
dark matter. Coupling model (III) includes also the case
$\Gamma<0$ which describes a transfer of energy from dark energy
to dark matter:
 \be
\Gamma~\left\{ \begin{array}{l} >0\\ <0 \end{array} \right. ~~
\Rightarrow ~~ \left\{ \begin{array}{l} \mbox{dark matter decays
$\to$ dark energy}\\ \mbox{energy transfer from dark energy $\to$
dark matter}\end{array} \right.
 \ee

It turns out that the resulting evolution equations do not allow a
two-dimensional representation of this model, since we cannot
eliminate $H$ from the energy balance equations~(\ref{cc}) and
(\ref{kg1}), using the variables $x(N),y(N)$. Equation~(\ref{reH})
must therefore be incorporated into the dynamical system. We do
this via a new variable $z$, chosen so as to maintain compactness
of the phase space:
\begin{equation}
  z = \frac{H_0}{H+H_0}\,.
\label{change2}
\end{equation}
Thus $0\leq z \leq 1$, and the compactified phase space now
corresponds to a cylinder of unit height and radius. We also
re-scale to a dimensionless coupling constant:
 \be
\gamma={ \Gamma \over H_0}\,.
 \ee
Then we arrive at the autonomous system
\begin{eqnarray}
  \label{x'}
x'&=&-3x+\lambda {\sqrt{6}\over 2}\,y^2+\frac{3}{2}x(1+x^2-y^2)
-\gamma\,\frac{(1-x^2-y^2)z}{2x(z-1)}\,,\\
  \label{y'}
y'&=&-\lambda {\sqrt{6}\over 2}\,xy+\frac{3}{2}y(1+x^2-y^2)\,,\\
  \label{z'}
z'&=& \frac{3}{2}z(1-z)(1+x^2-y^2)\,.
\end{eqnarray}
Since the system is invariant under $y \to -y$, the phase space
may be reduced to a unit semi-cylinder.

In order to determine the critical points and their stability, we
need to deal with the singularities at $x=0,z=1$ in
Eq.~(\ref{x'}). We rewrite the right-hand side of Eq.~(\ref{x'})
with a common denominator, in the form $x'=u(x,y,z)/[x(z-1)]$. Of
the total of seven points that give $y'=0=z'$ and $u=0$, three of
them also give $z=1$. The points A, B, C, D of Table~\ref{crit3}
are the critical points for the early universe, since $z
\rightarrow 0$ corresponds to $H \rightarrow \infty $. These are
the same critical points as occur in the uncoupled case. The new
points E, F, G with $z=1$ are late-universe versions of A, B, C
($z\to 1~\Rightarrow~ H\to 0$).

Our results are summarized in Table~\ref{crit3}. The stability
behaviour shown in Table~\ref{crit3} is based on the eigenvalues
of the linearized matrix, Eq.~(\ref{eigv}), which are shown in
Table~\ref{eigen1}. In fact, only the signs of the real and
imaginary parts of the eigenvalues are of importance. Therefore,
we consider the eigenvalue $\pm\infty$ simply as a positive/
negative eigenvalue.

\begin{table*}[!h]
\begin{center}
\begin{tabular}{|@{\hspace{0.05in}}c@{\hspace{0.05in}}|
@{\hspace{0.05in}}c@{\hspace{0.05in}}|
@{\hspace{0.05in}}c@{\hspace{0.05in}}|@{\hspace{0.05in}}
c@{\hspace{0.05in}}|@{\hspace{0.05in}}c@{\hspace{0.1in}}|
@{\hspace{0.05in}}c@{\hspace{0.05in}}|@{\hspace{0.05in}}
c@{\hspace{0.05in}}|
@{\hspace{0.05in}}c@{\hspace{0.05in}}|@{\hspace{0.05in}}
c@{\hspace{0.05in}}|} \hline&&&&&&&&\\[-0.3cm]
Point & $x_*$ & $y_*$ & $z_*$ & Stable? & $\Omega_{\varphi *}$  &
$w_{\text{tot} *}$ & Acceleration?
& Existence?\\[0.1cm]
\hline
\hline&&&&&&&&\\[-0.3cm]
A & 1 & 0 & 0 & saddle node for $\lambda>\sqrt{6}$& 1
&1 & no &  all $\lambda,\gamma$  \\
&    & & &unstable node for $\lambda<\sqrt{6}$&  & & &\\[0.2cm]
\hline&&&&&&&&\\[-0.3cm]
 B & -1 & 0 & 0
& unstable node for $\lambda>-\sqrt{6}$
 & 1 & 1 & no &  all $\lambda,\gamma$\\
& & &  &saddle node for $\lambda < -\sqrt{6}$ & & & &\\[0.2cm]
\hline&&&&&&&&\\[-0.3cm]
C & $\displaystyle\frac{\lambda}{\sqrt{6}}$ &
$\sqrt{1-\displaystyle\frac{\lambda^2}{6}}$& 0 &saddle node & 1 &
$\displaystyle\frac{\lambda^2}{3}-1$&
$\lambda ^2 <2$ &  $\lambda^2 \leq 6$\\[0.4cm]
\hline&&&&&&&&\\[-0.3cm]
D & $\displaystyle\frac{\sqrt{6}}{2\lambda}$ &
$\displaystyle\frac{\sqrt{6}}{2 \lambda}$ &0 &  saddle node for
$3<\lambda^2<\displaystyle{24 \over 7}$&
$\displaystyle\frac{3}{\lambda^2}$ &
 $0$ & no &  $\lambda^2 \geq 3$\\
&&&& saddle focus for $\lambda^2>\displaystyle{24 \over 7}$
&&&&\\[0.2cm]
\hline&&&&&&&&\\[-0.3cm]
E & 1 & 0 & 1 & stable node for $\lambda>\sqrt{6}$ and $\gamma >
0$ & 1 &1 & no &
 all $\lambda,\gamma$ \\
& & &
& saddle node for $\lambda>\sqrt{6}$ and $\gamma < 0$ & & & &\\
& & &
& saddle node for $\lambda<\sqrt{6}$ and all $\gamma$ & & & &
\\[0.2cm]
\hline&&&&&&&&\\[-0.3cm]
 F & -1 & 0 & 1
& saddle node for $\lambda>-\sqrt{6}$ and all $\gamma$ & 1 & 1 &
no &
all $\lambda,\gamma$\\
& & &
& stable node for $\lambda<-\sqrt{6}$ and $\gamma > 0$ & & & &\\
& & &
& saddle node for $\lambda<-\sqrt{6}$ and $\gamma < 0$ & & & &
\\[0.2cm]
\hline&&&&&&&&\\[-0.3cm]
G & $\displaystyle\frac{\lambda}{\sqrt{6}}$ &
$\sqrt{1-\displaystyle\frac{\lambda^2}{6}}$& 1 &stable node for
$\gamma > 0$ & 1 & $\displaystyle\frac{\lambda^2}{3}-1$
&  $\lambda ^2 <2$ &  $\lambda^2 \leq 6$\\
& & &
& saddle node for $\gamma < 0$ & & & &\\[0.4cm]
\hline
\end{tabular}
\end{center}
\caption[crit]{\label{crit3} The properties of the critical points
for the coupling model (III).}
\end{table*}

\begin{table*}[!h]
\begin{center}
\begin{tabular}{|c|c|c|c|c|} \hline
Point &$x_*$& $y_*$&$z_*$& Eigenvalues \\ [0.1cm] \hline
\hline &&&&\\[-0.3cm]
A & 1 & 0 & 0
& $3;\,\,3;\,\,3-{\displaystyle\frac{\sqrt{6}\lambda}{2}}$ \\[0.2cm]
\hline&&&&\\[-0.3cm]
B & -1 & 0 & 0 & $3;\,\,3;\,\,3+{\displaystyle\frac{\sqrt{6}
\lambda}{2}}$
 \\[0.2cm]
\hline&&&& \\[-0.3cm]
C& $\displaystyle\frac{\lambda}{\sqrt{6}}$ &
$\left(1-\displaystyle\frac{\lambda^2}{6}\right)^{1/2}$& 0 &
$\frac{\lambda^2}{2};\,\, \lambda^2-3;\,\,
{\displaystyle\frac{\lambda^2}{2}}-3 $ \\[0.2cm]
\hline
&&&&\\[-0.3cm]
D& $\displaystyle\frac{\sqrt{6}}{2\lambda}$ &
$\displaystyle\frac{\sqrt{6}}{2 \lambda}$ &0 &
${\displaystyle\frac{3}{2}}; \,\,
-{\displaystyle\frac{3}{4}}\left(\lambda \pm \sqrt{24-7
    \lambda^2}\right)$\\[0.2cm]
\hline&&&&\\[-0.3cm]
E& 1 & 0 & 1 &
$-3;\,\,3-{\displaystyle\frac{\sqrt{6}\lambda}{2}};\,\,\,
 -\sgn(\gamma)\,\infty$ \\[0.2cm]
\hline&&&&\\[-0.3cm]
F& -1 & 0 & 1 & $-3;\,\, 3+{\displaystyle\frac{
\sqrt{6}\lambda}{2}};\,\,\,
-\sgn(\gamma)\,\infty$ \\[0.2cm]
\hline &&&&\\[-0.3cm]
G& $\displaystyle\frac{\lambda}{\sqrt{6}}$ & $\left(
1-\displaystyle\frac{\lambda^2}{6}\right )^{1/2}$& 1 &
$-{\displaystyle\frac{\lambda^2}{2}};\,\,
{\displaystyle\frac{\lambda^2}{2}}-3;\,\, - \sgn(\gamma)\,\infty$
\\[0.2cm]
\hline
\end{tabular}
\end{center}
\caption[crit]{\label{eigen1} Critical points and associated
eigenvalues for coupling model (III). The $\infty$ appears due to
the limit $z \rightarrow 1$.}
\end{table*}

Apart from the extra dimension in its phase space, model (III)
differs from models (I) and (II) in one key aspect:
\begin{itemize}

\item
In models (I) and (II), the new behaviour introduced by coupling
includes the possibility of an accelerated scaling solution (point
D), characterized by Eq.~(\ref{scal}). In model (III), no such
accelerated scaling solution is possible. Instead, the accelerated
attractor introduced by the coupling is point G, which has
 \be \label{att3}
\Omega_{c *}=0\,,~~\Omega_{\varphi *}=1\,.
 \ee
This is the same qualitative behaviour as the standard
$\Lambda$CDM model.

\end{itemize}

\begin{figure*}[!ht]
\centering
\includegraphics[width=0.48\textwidth]{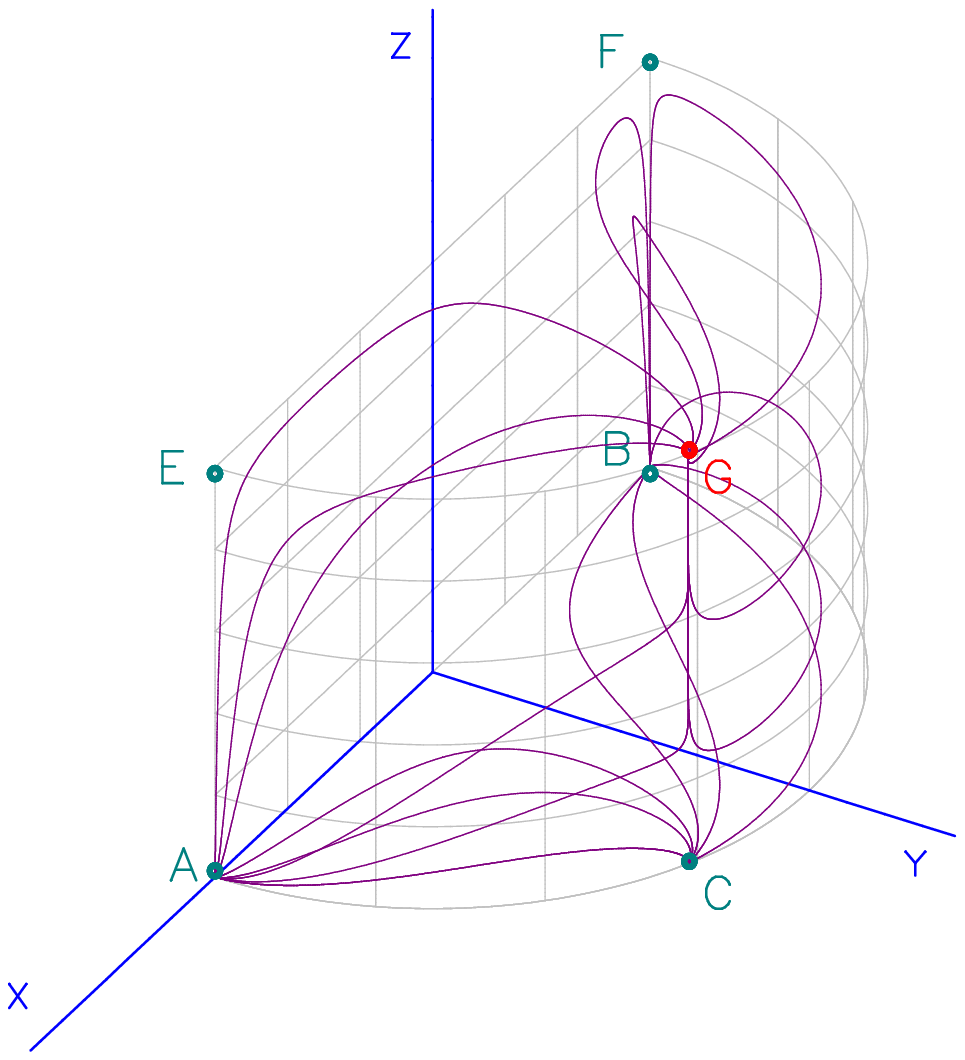}
\hfill
\includegraphics[width=0.48\textwidth]{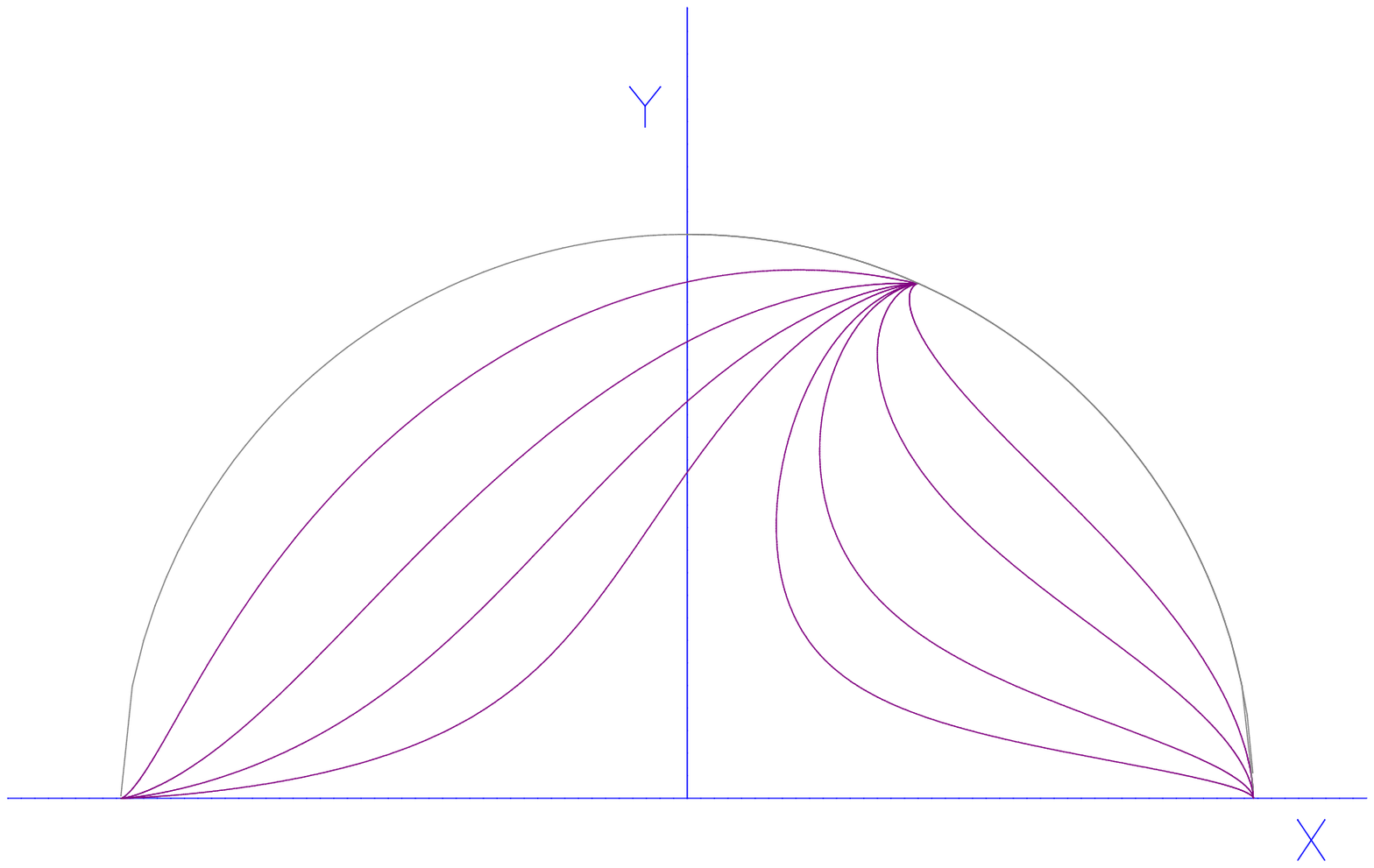}
\caption{Phase-space trajectories for coupling model (III), with
$\lambda=1$ and $\gamma=10^{-6}$. The right hand plot is the
projection. The global attractor G (see Table~\ref{crit3}) is
apparent.} \label{lambda_one}
\end{figure*}

\begin{figure}[!ht]
\centering
\includegraphics[width=0.48\textwidth]{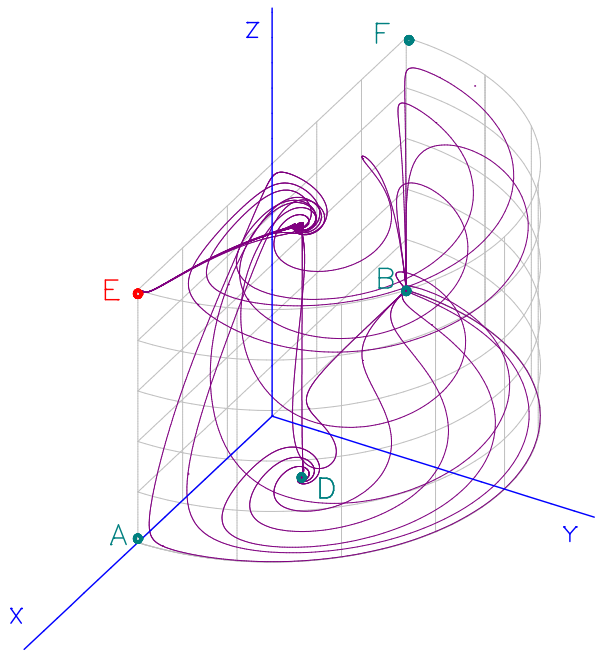}
\caption{\label{lambda_four} Phase-space trajectories for coupling
model (III), with $\lambda=4$ and $\gamma=10^{-6}$. This plot
shows the attractor E.}
\end{figure}

Our analytical results are supported by numerical integrations,
giving a consistent picture of the dynamical properties of
coupling model (III). Illustrative examples are shown in
Figs.~\ref{lambda_one} and \ref{lambda_four}. Trajectories for
$\lambda=1$ are shown in Fig.~\ref{lambda_one}, showing the
critical point G in Table~\ref{crit3}. In Fig.~\ref{lambda_four},
we plot trajectories when $\lambda=4$, and the stable node that
corresponds to the point E is apparent.

An interesting feature of the trajectories before reaching the
global attractor (E, F or G, depending on the values of $\lambda$
and $\gamma$), is that they seem to be focused near a point that
is vertically above the early-universe critical point D, and close
to the $z=1$ surface. Analysis of the system shows that the
coordinates and derivatives of this (non-critical) point are
 \bea
&& x={\lambda \over \sqrt{6}}\,,~y=\sqrt{1-{\lambda^2\over
6}}\,,~z=1-{\gamma \over \gamma+\lambda^2-3}\,,\\
&& x'=0\,,~~y'=0\,,~~ z'=O(\gamma)\,.
 \eea
In the limit of zero coupling, we have $z'\to 0$, and this point
collapses to the $z=0$ critical point D. The deviation of this
point from being critical is $O(\gamma)$. When $\gamma$ is small,
as in the plots, this explains the presence of focusing of
trajectories in the numerics.

\section{Conclusions}
\label{sec:concl}

We considered the background dynamics of a universe dominated by
dark energy (in the form of exponential quintessence) and cold
dark matter, where there is energy exchange in the dark sector, as
in Eqs.~(\ref{cc}) and (\ref{kg1}),
 \[
\dot \rho_c + 3H\rho_c=-Q=-\left[ \dot \rho_{\varphi} +
3H(1+w_{\varphi})\rho_{\varphi}\right].
 \]
For the previously introduced forms of $Q$, given in
Eqs.~(\ref{I}) and (\ref{II}),
 \[
\mbox{(I):  }~~ Q=\sqrt{ 2 /3}\, \kappa\,\beta
\rho_c\dot\varphi\,,~~~ \mbox{(II):  }~~  Q= \alpha H \rho_c\,,
 \]
the phase space remains two-dimensional, as in the uncoupled case
$Q=0$. We found the properties of the critical points for all
signs of $\lambda, \alpha, \beta$. The results are summarized in
Tables~\ref{crit} and \ref{crit2}, and slightly extend previous
work~\cite{Amendola:1999qq,Holden:1999hm,Billyard:2000bh} in the
case of pressure-free matter ($w_c=0$). In both models, critical
point D is the cosmologically relevant point, because for nonzero
coupling it includes accelerated scaling attractor solutions,
\[
0<\Omega_{c *}\,,\Omega_{\varphi *}<1 ~\mbox{  and  }~
w_{\text{tot} *} <-{1\over3}\,.
\]
The stability behaviour of critical point D was investigated
numerically, and is shown in Fig.~\ref{DI} for model (I) and
Fig.~\ref{DII} for model (II).

Our main results are for a new coupling model~\cite{val}, defined
in Eq.~(\ref{III}),
\[
\mbox{(III):  }~~Q=\Gamma \rho_c\,.
\]
This has a similar form to model (II), but is more physical since
the transfer rate $\Gamma$ is determined only by local properties
of the dark sector interaction at each event, and is not dependent
on the universal expansion rate. When $\Gamma>0$, this new model
has the same form as simple models for the decay of dark matter
particles to radiation~\cite{Cen:2000xv}, or to
quintessence~\cite{Ziaeepour:2003qs}, and for the decay of the
curvaton field into radiation~\cite{Malik:2002jb}.

Model (III) requires a three-dimensional phase space, since the
Hubble rate cannot be eliminated from the equations for $x',y'$.
This makes the dynamics more complicated than for models (I) and
(II). In particular, a new set of late-time critical points arises
in (III), and considerable analytical effort is required to
identify these points and determine their stability properties.
Our results are summarized in Tables~\ref{crit3} and \ref{eigen1}.
We performed numerical integrations of the dynamical system in
order to confirm the analytical results, and examples of these
integrations are shown in Figs.~\ref{lambda_one} and
\ref{lambda_four}.

The cosmologically relevant critical point is G, which allows for
an accelerated critical solution (when $\lambda^2<2$). However,
this is not a scaling solution, since
\[
\Omega_{c *}=0\,,~~\Omega_{\varphi *
}=1\,,
\]
which is similar to the asymptotic behaviour of the standard
$\Lambda$CDM model. This accelerated critical solution is an
attractor when $\gamma>0$, i.e., for the case when dark matter is
decaying to dark energy. Note that in model (II), the decaying
dark matter case, $\alpha>0$, does not lead to any accelerated
attractor (see Table~II). Model (III) with $\Gamma>0$ produces an
interesting class of models where dark matter decays to dark
energy -- so that the primordial universe may have no dark energy
-- and where this decay eventually leads to dark energy dominance,
independent of initial conditions (since there is an attractor).
Although such models do not solve the coincidence problem in the
standard way, they may provide a new approach to the broader
problem of explaining why dark energy dominates over dark matter
only late in the universe's evolution.

The background dynamics for coupling model (III) show new features
not present in the previously investigated models (I) and (II). In
order to confront this model with observations, the cosmological
perturbations with a dark sector coupling of form (III) need to be
investigated (see Ref.~\cite{val}).

\[ \]
{\bf Acknowledgements:}\\
We thank  Luis Ure\~na-L\'opez, Elisabetta Majerotto,  Luca Parisi, Israel Quir\'os,  Jussi
V\"aliviita and Shinji Tsujikawa for useful discussions. GCC is  supported by the
Programme Alban, the European Union Programme of High Level
Scholarships for Latin America, scholarship no.~E06D103604MX and
the Mexican National Council for Science and Technology, CONACYT,
scholarship no.~192680. RL is supported by the University of the
Basque Country through research grant GIU06/37, and by the Spanish
Ministry of Education and Culture through research grants
FIS2004-01626 and FIS2004-0374-E. The work of RM is supported by
STFC.

\end{document}